\documentclass[fleqn,usenatbib]{mnras}

\usepackage{newtxtext,newtxmath}
\usepackage[T1]{fontenc}
\usepackage{ae,aecompl}
\usepackage{graphicx}	% Including figure files
\usepackage{amsmath}	% Advanced maths commands
\usepackage{amssymb}	% Extra maths symbols
\usepackage{epsfig}
\usepackage{epstopdf}

\def\be{\begin{equation}}
\def\ee{\end{equation}}
\title[Modified initial power and TBTF]{Modified initial power spectrum and too big to fail problem}

\author[Hamed Kameli, Shant Baghram]{
Hamed Kameli %$^{1}$
~and Shant Baghram \thanks{baghram@sharif.edu}
\\
Department of Physics, Sharif University of
Technology, P.~O.~Box 11155-9161, Tehran, Iran\\\
}

\date{Accepted XXX. Received YYY; in original form ZZZ}

\pubyear{2019}

\begin{document}
\label{firstpage}
\pagerange{\pageref{firstpage}--\pageref{lastpage}}
\maketitle

% Abstract of the paper
\begin{abstract}
The galactic scale challenges of dark matter such as "missing satellite" problem and "too big to fail" problem are the main caveats of standard model of cosmology. These challenges could be solved either by implementing the complicated baryonic physics or it could be considered as an indication to a new physics beyond the standard model of cosmology. The modification of collisionless dark matter models or the standard initial conditions are two promising venues for study. In this work, we investigate the effects of the deviations from scale invariant initial curvature power spectrum on number density of dark matter halos. We develop the non-Markov extension of the excursion set theory to calculate the number density of dark matter substructures  and dark matter halo progenitor mass distribution. We show that the plausible solution to "too big to fail" problem could be obtained by a Gaussian excess in initial power in the scales of $k_* \sim 3 \text{h/Mpc}$ that is related to the mass scale of $M_* \sim 10^{11} M_{\odot}$. We show that this deviation leads to the decrement of dark matter sub-halos in galactic scale, which is consistent with the current status of the non-linear power spectrum.  Our proposal also has a prediction that the number density of Milky way type galaxies must be higher than the standard case.
\end{abstract}

% Select between one and six entries from the list of approved keywords.
% Don't make up new ones.

%%%%%%%%%%%%%%%%%%%%%%%%%%%%%%%%%%%%%%%%%%%%%%%%%%

%%%%%%%%%%%%%%%%% BODY OF PAPER %%%%%%%%%%%%%%%%%%

\section{Introduction}
\label{Sec:1}
The standard model of cosmology is well established  with the observations of the cosmic microwave background (CMB) \cite{Aghanim:2018eyx} and the late time large scale structure (LSS) \cite{Alam:2016hwk}.
Despite the many successes of the standard model, there are couple of challenges for the vanilla model, such as the Hubble constant tension \cite{Freedman:2017yms} the $\sigma_8$ tension \cite{Kohlinger:2017sxk} and galactic scale challenges of dark matter \cite{Bullock:2017xww}. Each of these tensions can help us to understand the standard model of cosmology $\Lambda$CDM better or can lead to a dramatic change in our understanding of the Universe.
The galactic scale challenges, which are the main focus of this work, are mainly categorized as core-cusp problem, missing satellite problem \cite{Klypin:1999uc,Moore:1999nt} and Too-Big-To-Fail (TBTF) problem \cite{BoylanKolchin:2011de}. These challenges are all appeared in scales smaller than $\sim 1 Mpc$ and in mass ranges smaller than $\sim 10^{11}M_{\odot}$. {{These tensions are appeared in the comparison of the observational data with N-body simulation designed to probe the dark matter halos with the standard $\Lambda$CDM model \cite{Springel:2008cc,Wang:2012sv,Vogelsberger:2014dza}}}
The study of the local group small galaxies which can be done by a great detail with upcoming future surveys is an opportunity to address the small scale challenges of cold dark matter (CDM) paradigm  and test the fundamental theories in cosmology. In CDM scenario, we anticipate to have a cusp-like density profile, many galactic satellites and more massive satellites in the host halo of the Milky Way and Andromeda. However, the observations contradict with standard predictions. The small scale challenges of dark matter introduce a new arena for the birth of the alternative models of dark matter beyond the collisionless CDM paradigm. As an example to the alternatives of CDM , we can mention the Warm Dark Matter (WDM) models \cite{Bode:2000gq}, Self Interacting Dark Matters (SIDM) models \cite{Tulin:2017ara} and Fuzzy Dark Matter (FDM) models \cite{Hui:2016ltb,Maleki:2019xya}. Also these challenges bring new observational ideas to find the dark matter sub-halos in small scales \cite{Baghram:2011is,Erickcek:2010fc,Rahvar:2013xya,Asadi:2017ddk}.
It should be noted that many baryonic physics solutions are suggested for solving the missing satellite problem and the core-cusp problem \cite{Wetzel:2016wro}.\\
In this work, we introduce a novel point of view to address the galactic scale challenges of CDM, such as TBTF problem and missing satellite problem. We assert that the modification to the standard picture of early universe initial condition can be a plausible solution. In the standard inflationary paradigm, we  assume a nearly scale invariant, adiabatic, isotropic and nearly Gaussian statistics for the curvature perturbations \cite{Baumann:2009ds}. Any deviation from the standard characteristics which is mentioned above can be considered as a new window to investigate the physics of the early universe. We show the modification of initial condition (I.C.) would affect the late time LSS observations. The idea of using the large scale structure observations as a probe of early universe physics is studied vastly in literature \cite{Elgaroy:2001wu,Baghram:2013lxa,Baghram:2014nha,Namjoo:2014nra,Hassani:2015zat,Fard:2017oex}. It should be noted that there are alternative approaches such as the running of the spectral index \cite{Garrison-Kimmel:2014kia,Leo:2017wxg} or suppression of matter power spectrum in small scales \cite{Nakama:2017ohe}.\\
{{The main idea of this work is to introduce a deviation from the standard scale invariant power spectrum to reconcile a tension in late time universe in small scales. We should note that the initial power spectrum corresponding to the wavenumbers related to scales of ($\sim 10^4$ Mpc), almost the horizon
size today, down to ($\sim 10$ Mpc), corresponding to angular resolution of Planck satellite, is well constraint by CMB temperature fluctuation angular power spectrum \cite{Aghanim:2018eyx}. Also, it must be noted that the LSS observations such as weak gravitational lensing \cite{Camacho:2018mel}, galaxy distribution \cite{Percival:2009xn} and the Ly-alpha observation constrain the power spectrum even to smaller scales up to $k \simeq 1 \text{h/Mpc}$. \\
However, going further to more smaller scales to the wavenumbers $k \gtrsim 1-10   ~ \text{h/Mpc}$ (sub-galactic scales) and even further, the initial power spectrum  is not constrained  by cosmological or astrophysical observation. This situation is true for smaller scale (up to $k\sim 10^{23}\text{h/Mpc}$ or more) there is not well and strict observational constraints. However, there are some theoretical ideas to study the small scale power spectrum, from non-detection of  ultra-compact mini halos (UCMH) \cite{Bringmann:2011ut}, primordial black holes (PBH) \cite{Emami:2017fiy} and evaporation of PBH \cite{Dalianis:2018ymb}. We should mention  the ideas of using gravitational wave (non)detection in small scales \cite{Inomata:2018epa}and also  the observation of the CMB spectral distortion \cite{Sarkar:2017vls} which are used to study sub-galactic scale of initial power spectrum. \\
The situation works as a hint for us that there could be a possibility for deviation from scale invariance of power spectrum in smaller scales which will  have some effects on distribution of matter in sub-galactic scales. Accordingly, we propose to add a Gaussian power excess to initial curvature power spectrum in scale corresponding to $k \sim 3\text{h/Mpc}$ which is related to the mass scale of $10^{11} M_{\odot}$.
We use a Gaussian shape toy model for initial curvature power spectrum. Although, this power excess could be a generic feature of inflationary models. Actually, many inflation models predict excess power in small scales. For instance, deviation from invariant initial power could be made through different inflationary potential models \cite{Salopek:1988qh,starobinsky:1992spe}, multiple inflationary field \cite{Adams:1997de,Hunt:2004vt}, particle production in inflationary era \cite{Chung:1999ve,Barnaby:2009dd}, supersymmetric inflation model \cite{Randall:1995dj}, effect of super Planck scale physics \cite{Martin:2000xs}, primordial non-Gaussianity \cite{Bartolo:2004if,Chen:2010xka} and etc.}} \\
We use the non-Markov extension of the Excursion set theory (EST) \cite{Bond:1990iw,Zentner:2006vw,Nikakhtar:2016bju} to calculate number density of dark matter halos in small scales. The non-Markov extension of EST is developed to count number density of dark matter halos with more precision in small scales \cite{Nikakhtar:2018qqg}.
The structure of this work is as follow: In Sec. (\ref{Sec:2}), we introduce the TBTF problem and review the idea of the EST with its recent developments in calculating the number density of dark matter halos. Furthermore, we study the connection of late time observables to  early Universe physics. In Sec.(\ref{Sec:3}), we review the non-Markov extension of EST, which is an important theoretical ingredient for dark matter halo counts. In Sec.(\ref{Sec:4}), we study the effect of deviations in primordial power spectrum on number density, mass fraction in late time and propose our main results. In Sec.(\ref{Sec:5}), we have the conclusion and future remarks. \\
We set the cosmological parameters due to Planck results \cite{Ade:2015xua}: present time matter density parameter  $\Omega_m = 0.319$, the Hubble parameter $H_0 = 67.74  \text{km/s/Mpc}$, the spectral index of primordial curvature power spectrum $n_s = 0.967$, the amplitude of the initial scalar curvature power spectrum $\ln(10^{10}A_s) = 3.04 $.

%*****************************************************
%*****************************************************
%*****************************************************
\section{ Theoretical background: Galactic scale challenges and structure formation from early to late times}
\label{Sec:2}
In this section, we discuss the theoretical backgrounds for this work. In first subsection, we briefly go through the galactic scale challenges of CDM, specially the TBTF problem. In second subsection, we introduce the EST and the relation between initial condition and late time power spectrum of the matter. In third subsection we study the non-linear matter power spectrum.
% ********************
% ********************
\subsection{Missing satellite and TBTF problem}
The standard CDM paradigm predicts that the dark matter halos must be present in all scales up  to the kinetic decoupling mass scale \cite{Profumo:2006bv}, accordingly in the mass range of $\sim 10^7 M_{\odot}$ we anticipate to have thousands of satellites. However, the observed dwarf galaxies are in order of $\sim 50$. Despite the missing satellite problem which has many proposed baryonic solutions \cite{Brooks:2012ah}, there is a more severe problem.
In a series of nominal works by Boylan-Kolchin et al. \cite{BoylanKolchin:2011de} (see more references in \cite{Bullock:2017xww} ), they showed that the population of the Milky way's brightest dwarf Spheroidal (dSph) galaxies do not match the prediction of CDM based simulations.
The classic dwarf galaxies which are mainly dark matter dominated % which are listed in Table{\ref{tab:sat}}
have the most luminous stellar distribution and velocities. With respect to the N-body simulations, we anticipate that the classic dSph should populate the most massive dark matter sub-halos. But in contrary to what we expect, the observations show low circular velocities.
More precisely, simulations predict $\sim O(10)$ sub-halos with maximum circular velocity  $V_{c(max)} > 30$ km/s, where the bright Milky Way dSphs have stellar velocities corresponding to the sub-halos with less velocity (e.g. $12 \text{km/s}<V_{c(max)}<25$ km/s). It is exciting that we notice the same problem in satellites of our neighbor galaxy Andromeda  \cite{Tollerud:2014zha}. Now the question is how it comes that the most massive sub-halos fail to have baryonic counterparts. Due to their deep potential wells, it comes almost unlikely that baryonic feedbacks can prevent the gas accretion and suppress the galaxy formation. Hence, these substructures are too big to fail to form stars! This is the reason that we called the problem TBTF.

%\begin{table}
%	\begin{tabular}{ |c|c|c| }
%		\hline
%		& Satellite & Absolute Visual Mag \\ \hline
%		%\multirow{4}{*}{\rotatebox[origin=c]{90}{ CMB+BAO+R16 }} & $\chi^2=7.84$ & $\chi^2=4.93$ \\
%		& Large Magellanic Cloud &  1 \\
%		& Small Magellanic Cloud &  1 \\
%		& Fornax  &  1  \\
%		& Sculptor &  1 \\
%		& Sagittarius &  1  \\
%		& Draco &  1 \\
%		%\multirow{3}{*}{\rotatebox[origin=c]{90}{ + Lyman-$\alpha$ BAO }} & $\chi^2=19.25$ & $\chi^2=17.11$ \\
%		& Ursa Major I &  1  \\
%		& Carina &  1  \\
%		& Sextan &  1  \\
%		\hline
%	\end{tabular}
%	\caption{\label{tab:sat}The list of classic satellites of Milky Way}
%\end{table}
An interesting point is that the TBTF problem is observed in the field dwarf galaxies \cite{Papastergis:2014aba}. This is important because all the mechanism of baryon extraction from a dwarf galaxy via tidal interactions are not applicable in field galaxies. The root of these challenges is that we assume that the dark matter halos must exist in all the mass ranges we discussed. The main proposal of this work is related to this point. We anticipate a modified initial condition can change the distribution of dark matter halos in a sense that it reconciles the galactic scale problems in CDM. %In the next subsection, we discuss the large scale structure formation, the linear power spectrum and its connection with the number density of the structures.

% ********************
% ********************
\subsection{Large scale structure: from matter power spectrum to number density of dark matter halos}
\begin{figure}
	\includegraphics[width=\columnwidth]{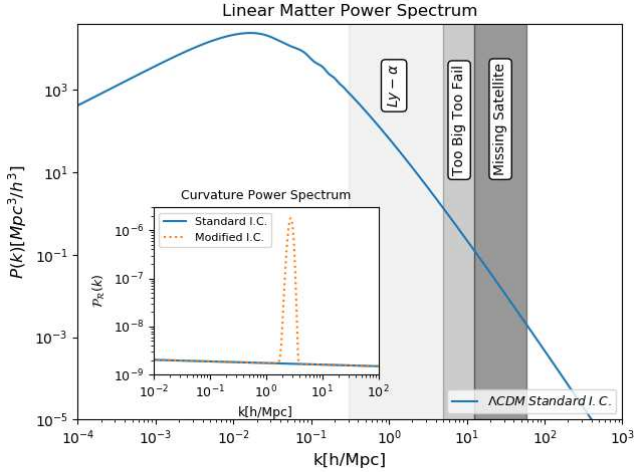}
    \caption{The linear matter power spectrum  is plotted versus the wavenumber.  The grey regions show the span of the wavenumbers of Ly-$\alpha$ observations, TBTF problem and missing satellite one. In the inset figure, we show the primordial curvature dimensionless power spectrum versus wavenumber. The modified Gaussian bump-like power is shown with dotted  line.} \label{fig-ps}
\end{figure}

The idea of Press and Schechter (PS) \cite{Press:1973iz} is to relate the number density of dark matter halos in non-linear regime to the linear matter power spectrum. The PS idea is reexpressed in a new perspective under the framework of Excursion Set Theory (EST) \cite{Bond:1990iw}. In EST a two dimensional space of density contrast versus the smoothing scale (the variance correspondingly) is used to find the number density of structures via the process of barrier crossing. The barrier in density contrast $\delta_c\simeq 1.69$ is the linear mapping of spherical collapse density to the present time. Accordingly, the number density of dark matter halos $n(M)$ in the mass range of $M$ and $M+dM$ is as below
\begin{equation}
n(M)dM= \frac{\rho_{ \scriptscriptstyle m}}{M}f_{ \scriptscriptstyle FU}(S)|\frac{dS}{dM}|dM ,
\end{equation}
where $\rho_{\scriptscriptstyle m}$ is the  matter density of the Universe. $S$ is the variance of perturbations in linear scale and $f_{\scriptscriptstyle FU}$ is the fraction of trajectories which they have their first up-cross in the variance range of $S$ and $S+dS$. The variance in redshift $z=0$ is related to linear regime matter power spectrum as
\begin{equation} \label{eq:variance}
S=\sigma^2(R)=\frac{1}{2\pi^2}\int dk k^2P_m(k,z=0)\tilde{W}^2(k;R),
\end{equation}
where $P_m(k,z=0) = |\delta_m(k,z=0)|^2$ is the linear matter power spectrum in present time ($\delta(k,z)$ is the matter density contrast in Fourier space), $\tilde{W}(k;R)$ is the Fourier transform of the smoothing function with radius $R$ in real space. In Figure (\ref{fig-ps}), we plot the curvature and matter power spectrum versus the  wavenumber. Moreover, we show the deviation in initial condition (I.C.) in the inset figure. As it is emphasized in the introduction, the main idea of this work is to modify the initial condition, where we can address the galactic scale problems.  We should note that in the wavenumbers of $k<0.2 \text{h/Mpc}$  the linear matter power spectrum is introduced  by the Poisson equation and evolution of potentials in the cosmological background. The grey regions labeled with Lyman-$\alpha$ (Ly-$\alpha$) is related to the non-linear matter power spectrum, which is well constraints with the Ly-$\alpha$ absorbtion line power spectrum. The smaller scales correspond to the galactic scales mass ranges and scales where the TBTF and missing satellite problem emerged. These scales are not constrained by cosmological observations. Accordingly there is a room for probing the modified powers in this regime.  The modified initial condition can change the power spectrum, a schematic graph is shown with  dotted lines. In Section \ref{Sec:4} we discuss the modification in more detail.\\ \cite{Bond:1990iw} show  by choosing a k-space sharp filter, the walks in the 2D plane of linear density contrast-variance will execute a Markov random walk. In this case, the first up-crossing fraction is determined analytically by absorbing barrier solutions as
\be
f_{FU}(S,\delta_c(z) dS = \frac{1}{\sqrt{2\pi}} \frac{\delta_c(z)}{S^{3/2}}\exp[-\frac{\delta^2_c(z)}{2S}]dS,
\ee
where the redshift dependence comes from the linear barrier (i.e. $\delta_c(z) = \delta_c D(z=0) / D(z)$, where $D(z)$ is the growth function). In the upper panel of Figure (\ref{fig:traj}), we plot the Markov trajectories of linear density versus variance. It is obvious in the plot that the trajectories are very jagged. The k-space sharp filter is not a physical choice for the smoothing scale. In Section \ref{Sec:3}, we discuss the non-Markov version of EST to choose a more realistic smoothing function.\\
In the context of EST, we can also calculate the merger history of dark matter halos. For this task, the distribution of dark matter progenitors can be find by  counting multiple up-crossings. In the Markov case, which is an obvious extension of the first up-crossing distribution we have
\be \label{eq:condition}
f_{FU}(S_1,\delta_1|S_2,\delta_2) = \frac{1}{\sqrt{2\pi}}\frac{\delta_1 - \delta_2}{(S_1 - S_2)^{3/2}}\exp[\frac{-(\delta_1 - \delta_2)^2}{2(S_1 - S_2)}],
\ee
where $S_1$ is related to a mass $M_1$ which is merged to a halo with larger mass $M_2$ corresponding to the variance $S_2$. The smaller(larger) halo is observed in redshift $z_1$($z_2$). The redshift dependence is encapsulated in the barriers via the growth function. The smaller (larger) halo has a barrier $\delta_1 = \delta_c/D(z_1)$ ($\delta_2=\delta_c / D(z_2)$ ). Note that the conditional number density of structures in the mass range of $M_1$ and $M_1+dM_1$ can be found by $n(M_1,t_1|M_2,t_2)dM_1 = (M_2/M_1)\times f_{FU}(S_1,\delta_1|S_2,\delta_2) \times |{dS_1}/{dM_1}|dM_1$. In the next section, we describe that how we can find this conditional probabilities in non-Markov case.\\
Now we have to relate the late time matter power spectrum to the initial curvature perturbation power spectrum via the Poisson equation $k^2 \Phi(k,z) = 4\pi G \rho_m(z) (1+z)^{-2}\delta_m(k,z)$ and the evolution of the gravitational potential $\Phi(k,z)$ in the different epochs of cosmology. Accordingly the density contrast can be written as
\be
\delta(k,z) = \frac{2}{3} \frac{k^2 \Phi (k,z) / (1+z)}{\Omega_m H_0^2}.
\ee
Now the gravitational potential can be related to the initial value of potential as $\Phi(k,z) = \frac{9}{10}T(k) D(z) (1+z) \Phi_{ini}$, where $T(k)$ is the transfer function, (we use the Eisenstein-Hu transfer function \cite{Eisenstein:1997ik}) and $\Phi_{ini}$ is the initial value of potential related to the curvature perturbation ${\cal{R}}_k$ as $\Phi_{ini} = \frac{2}{3} {\cal{R}}_k$.
The curvature perturbation in the standard model of cosmology with nearly scale invariant initial conditions is parameterized as
\be
{\cal{P}}_{\cal{R}} (k) = \frac{1}{2\pi^2}k^3 P_{\cal{R}}(k)=A_s (\frac{k}{k_p})^{n_s -1},
\ee
where $A_s$ is the amplitude of primordial power, $n_s$ is the spectral index of perturbations and $k_p = 0.002 Mpc^{-1}$ is the pivot wavenumber.
Summing all this evolution, we can write the linear power spectrum as below
\be
P_m(k,z) = A_l T^2(k)D^2(z)k^{n_s},
\ee
where $A_l$ is the amplitude of the linear power spectrum related to the initial amplitude $A_l = (8\pi^2 / 25) k^{1-n_s}_p \Omega^{-2}_m H^{-4}_0 A_s$. So any deviation from the standard initial power spectrum of curvature is transferred to the matter power spectrum. This is an approximation, assuming that the modified initial condition does not change the evolution of gravitational potential in the course of cosmic expansion and the transfer function remained the same.% In the next subsection, we discuss the non-linear power spectrum which is essential to check the consistency of our proposed idea with the observational data in the scales mildly larger that the TBTF scale.
% ********************
% ********************
\subsection{Non linear matter power spectrum}
The non-linear evolution of the perturbation enhance the matter power spectrum in small scales, this enhancement is in agreement with the observational data, such as matter power spectrum in small scales. The non-linear power spectrum can be modeled by halo model of dark matter structures \cite{Cooray:2002dia}.
The power spectrum in small scales can be decomposed into two halo $P^{2h}(k)$ and one halo $P^{1h}(k)$ terms as
\be \label{eq:1h2h}
P(k)=P^{1h}(k)+P^{2h}(k) ,
\ee
where the one and two halo terms are defined as below respectively,
\be
P^{1h}(k)=\int dm n(m) (\frac{m}{\bar{\rho}})^2 |u(k|m)|^2,
\ee
%\begin{eqnarray}
\begin{flalign}
P^{2h}(k) &= \int dm_1 n(m_1)(\frac{m_1}{\bar{\rho}})u(k|m_1) \\ \nonumber
&\times   \int dm_2 n(m_2) (\frac{m_2}{\bar{\rho}})u(k|m_2)P_{hh}(k|m_1,m_2),
\end{flalign}
%\end{eqnarray}
where $n(m)$ is the number density of dark matter halos with mass $m$ discussed in previous section and $u(k|m)$ is the Fourier transform of dark matter distribution in a halo defined as
\be
u(k|m)= \int_0^{r_{vir}} dr 4\pi r^2 \frac{\sin kr}{kr} \frac{\rho(r|m)}{m}.
\ee
We use the Navarro-Frenk-White (NFW) profile for dark matter halos \cite{Navarro:1996gj}.
The $P_{hh}(k|m_1,m_2)$ is the halo-halo power spectrum which is related to the linear scale matter power as below:
\be
P_{hh}(k|m_1,m_2)\simeq b_1(m_1)b_2(m_2)P_m(k),
\ee
where $b_{1(2)}$ are the dark matter halo bias which has a mass dependency $m_{1(2)}$ appeared in equation \ref{eq:1h2h}. The halos bias is discussed vastly in literature \cite{Mo:1995cs,Sheth:1999mn,Tinker:2010my}. We assume a simple bias obtained from EST and peak-background splitting as $b(m,z) = 1+ (\nu^2(z) -1) / \delta_c(z)$, where $\nu \equiv \delta_c (z)/ S^{1/2}$ is the height parameter \cite{Mo:1995cs}.

%*****************************************************
%*****************************************************
%*****************************************************
\section{non-Markov extension of EST}
\label{Sec:3}
\begin{figure}
	\includegraphics[width=\columnwidth]{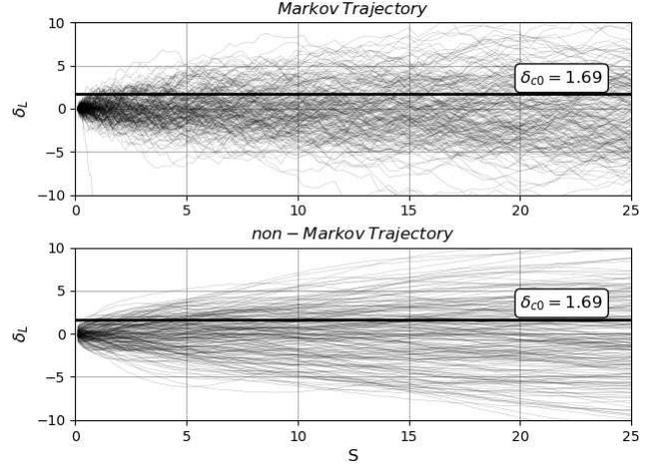}
    \caption{The trajectories of linear density contrast versus variance is plotted for two cases of Markov and non-Markov walks. The non-Markov trajectories are produced by Cholesky method and with the linear matter power spectrum of the standard model of cosmology.} \label{fig:traj}
\end{figure}
The idea of using the initial box in the early universe, where the perturbations are linear and Gaussian, introduce a very interesting arena to study the relation of the number density of non-linear objects and the physics of initial conditions. However the original solutions proposed by Bond et al. \cite{Bond:1990iw} assumes a specific window function, known as the k-space sharp filter which results to an analytical solution for the mass profile of dark matter halo. The more realistic smoothing functions (i.e. Gaussian or top-hat in real space), make the correlated trajectories in the EST plane. It means that the walks have memory and they will be a non-Markov process.  To solve the first up-crossing (FU) problem in 2D plane of the EST and to avoid the cloud in cloud problem, many proposals have been discussed in the literature(see the introduction of \cite{Nikakhtar:2018qqg}). A very sophisticated idea is to use the up-crossing approximation instead of the FU for large dark matter halos \cite{Musso:2012qk,Musso:2013pha}. However, this approximation breaks down in small mass ranges.
In Nikakhtar et al. \cite{Nikakhtar:2018qqg}, beside the analytical approximations known as Hertz and Stratonovich approximations, a numerical exact method known as the Cholesky decomposition have been introduced to make the trajectories. For this method, we assume that $\delta_{R}(x)$ is the linearly evolved  density contrast in position $x$ with a smoothing scale $R$. In $\Lambda$CDM the smoothing scale is a monotonic function of variance $S$ introduced in equation (\ref{eq:variance}).% In EST, one can find the linearly evolved density contrast in terms of the smoothing scale(variance) which resembles a stochastic process.
And also for each position in the Universe, one can find the corresponding trajectory.  The Cholesky method is a numerical framework to make the trajectories with the correct ensemble (statistical) properties. The statistical properties of walks depends on the correlation between the steps and the nature of smoothing function. In order to quantify this dependency we introduce the correlation of the heights of the walks in two different scale as
\be
\langle \delta_i\delta_j\rangle\equiv C_{ij} = \int \frac{dk}{k}\frac{k^3P(k)}{2\pi^2}\tilde{W}(kR_i)\tilde{W}(kR_j),
\ee
where $P_m(k)$ is the linear matter power spectrum and $\tilde{W}(kR_i) [\tilde{W}(kR_j)]$ is the window function in the smoothing scale $R_i  [R_j]$ respectively. Note that $C_{ii}=S_i$ is the variance of $\delta_R$ when the smoothing scale is $R_i$. In the Cholesky method, the height of the walk in a step $n$ defined as
\be
\delta_{n}=\langle \delta_n | \delta_{n-1}, ... ,\delta_1 \rangle + \sigma_{n|n-1, ..., 1}\xi_n ,
\ee
where by the first term  we mean that the height in $n$th step is  a linear function of the previous walks and second term depends only to the scales of smoothing window (or the corresponding variance $S_1,...,S_n$). The $\xi_n$ is zero mean unit variance Gaussian random number (with the ensemble property of $\langle \xi_n\xi_m = \delta_{nm} \rangle$). The distribution of density contrast is a multivariate Gaussian. In the case, if the smoothing window function is a k-space sharp filter, then $C(S,S')=min(S,S')$, which recovers the Markov walks and the original EST formulation is applicable. However, if we use  more physically  meaningful window function like real space top hat $\tilde{W}(x) = 3\frac{j_1(x)}{x}$ or the Gaussian window function $\tilde{W}(x) = \exp[-x^2/2]$ (note that $x=kR$) , we will come up with a non-Markov more smoother walks. In Figure (\ref{fig:traj}) lower panel, we plot the ensemble of trajectories for standard model of cosmology with a "Gaussian" window function. This is the first ever obtained plot of trajectories for a $\Lambda$CDM cosmology with the EH transfer function \cite{Eisenstein:1997ik}. For the future reference, also the ensemble of trajectories for further investigation with modified initial conditions is obtained.
In the plane of the trajectories, there is a barrier $\delta_c$ which is related to the collapse model in non-linear regime, which could be a scale dependent quantity as well \cite{Sheth:2001dp}. In this work we assume a constant barrier $\delta_c$, which is an approximation in order to give the general view of how modified initial condition can change the distribution of the matter in late time.
%In Figure(\ref{fig:fig2-traj}), we plot the Markov and non-Markov trajectories of a linear density contrast versus variance in the context of EST. We should note that the non-Markov ones are produced by the Cholesky method described in much detail in below, using the linear matter power spectrum of $\Lambda$CDM
We should note that the generation of the exact trajectories is very important in small scales. This is because in EST  approach the first up-crossing of trajectories from a specified barrier is related to the non-linear object with the mass $M$ which itself is related to the smoothing scale $R$ (interchangeably the variance $S$). Accordingly, the more precise method is essential for number count problem for low mass dark matter halos. \\
Coming back to the Cholesky method, in order to produce the trajectories, we note that covariance matrix $C_{ij}$ is a real, symmetric and positive definite. Accordingly this matrix can be decomposed in a unique way $C={\bf{L}}{\bf{L}}^T$, in which ${\bf{L}}$ is a lower triangular matrix. Now we can use the ${\bf{L}}$ to generate the ensemble of trajectories.
The density contrast in scale $R_i$ is obtained as
\be
\delta_i = \sum_{j} {\bf{L}}_{ij} \xi_j,
\ee
where $\xi_j$ is random number with Gaussian distribution. In this case, $\delta_i$ will have the correct correlation between heights which is given by
\be
\langle \delta_i \delta_j \rangle = \sum_{m,n} L_{im} L_{jn} \langle \xi_m \xi_n \rangle = {\bf{L}}{\bf{L}}^T = C.
\ee
The procedure of Cholesky decomposition is described in appendix A of \cite{Nikakhtar:2018qqg}.
It is interesting to note that by making the trajectories with a given matter power spectrum and smoothing scale, we can find the conditional mass function (equation \ref{eq:condition}), by just counting the multiple up-crossing in different scales and barriers.
Now in the next section, we introduce the modified initial condition and we will discuss our results.

%*****************************************************
%*****************************************************
%*****************************************************
%*****************************************************

\section{Results: Modified initial condition and non-linear structure formation}
\label{Sec:4}
In this section, in order to solve the TBTF and missing satellite problem we introduce a toy model. The model is a modification in the initial matter power spectrum  to examine the effect of initial conditions on the late time observables. {{ The power excess in initial condition could be a generic feature of inflationary models. For example, axion monodromy models naturally introduce a bump-like feature in power spectrum which is analyzed by Planck collaboration as well \cite{Akrami:2018odb}. ( For models with the idea of modified initial power spectrum see \cite{Salopek:1988qh,starobinsky:1992spe,Adams:1997de,Hunt:2004vt,Chung:1999ve,Barnaby:2009dd,Randall:1995dj,Martin:2000xs}). }}\\
The modification to initial power spectrum is defined as
\be
{\cal{P}}_{\cal{R}}(k)= \bar{{\cal{P}}}_{\cal{R}}(k)+{\cal{P}}^{bump}_{\cal{R}}(k),
\ee
where ${\cal{P}}^{bump}_{\cal{R}}(k)$, is Gaussian function which is parameterized as
\be
{\cal{P}}^{bump}_{\cal{R}}(k) = \frac{A_{b}}{\sqrt{2\pi} \sigma_b}\exp[-(k-k_*)^2/2\sigma^2_b],
\ee
where $A_b$ is the amplitude of Gaussian modification, the $k_*$ is the specific wavenumber which the bump is applied and $\sigma_b$ is variance of the bump. This modification introduces a deviation in primordial matter power spectrum which is plotted in Figure (\ref{fig-ps}), with $A_b=2000$, $k_* =2.72 h/Mpc$ (related to mass scale of $M_* = 10^{11} M_{\odot}$ )and $\sigma_b=0.25$. (Note: This specific numbers are chosen to decrease the number of dark matter halos in the TBTF scale and also be consistent with non-liner power spectrum constraints).\\

\begin{figure}
	\includegraphics[width=\columnwidth]{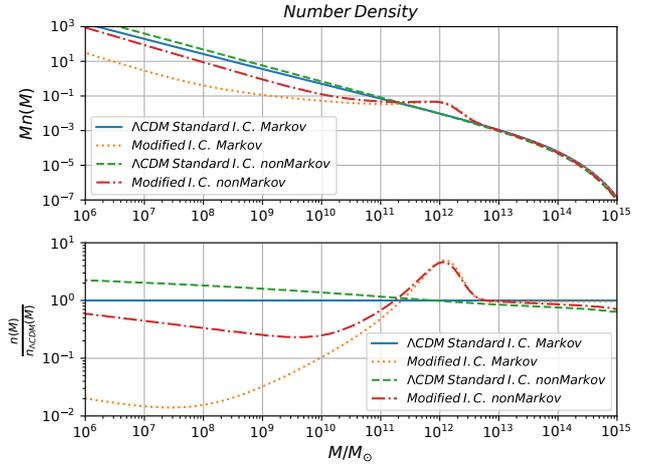}
    \caption{Upper panel: the number density of dark matter halos for the Markov and non-Markov cases of standard and modified I.C. models are plotted.
lower panel: ratio of mentioned number density with respect to the standard Markov number density is plotted. } \label{fig:nM}
\end{figure}

\begin{figure}  \label{fig:fig4-cond12}
	\includegraphics[width=\columnwidth]{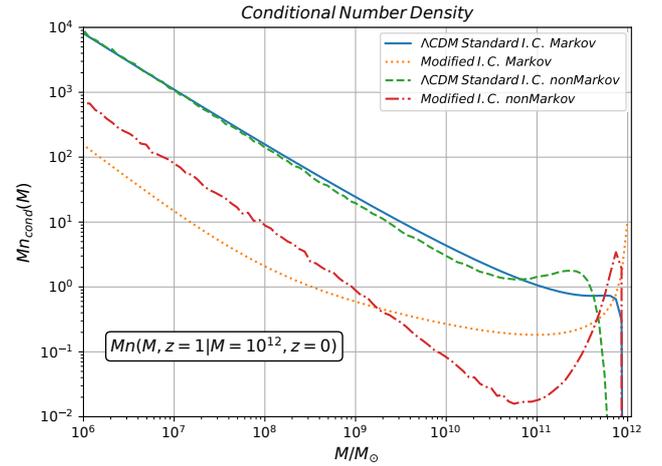}
    \caption{The conditional mass function of dark matter sub-halos in $z=1$ which must be hosted by a milky way like dark matter halo.} \label{fig:cond}
\end{figure}

A simple intuition is that the excess in matter power spectrum increase the number density of dark matter halos. However, the hierarchical structure formation is more complicated. The specific  mass scale $M_*$ which is related to specific wavenumber $k_*$  remains almost unchanged. The dark matter halos smaller than the specific mass ($M<M_*$) suppressed by the process of merging to the larger dark matter halos $M>M_*$. {{ This process, give us the opportunity to solve the TBTF problem by this proposal:\\ {\it{The modified initial power spectrum by an excess in a specific scale (i.e. $k_* \simeq 2.72 \text{h/Mpc}$), which is much catchable in inflationary models than the suppression in power, decrease the dark matter halo number in the scales $M<M_* =10^{11}M_{\odot}$, which brings up a new idea for the deficit of dark matter halos in galactic scales.}}\\
We should note that this idea does not use the baryonic physics to address the TBTF problem, instead we assert that the solution to this problem is due to deficit of the number density of dark matter halos in larger mass range which host the most luminous satellites of the main halo.}}
However, we should be cautious that this modification causes an increase in dark matter halos in larger masses mildly larger than $M \geq 10^{11}M_{\odot}$ and this means that we have to be consistent with the observations in non-linear scale.
In Figure (\ref{fig:nM}), we plot the number density of dark matter halos versus mass.  In this figure we have the standard Markov prediction of $\Lambda$CDM. We also plot the number density of dark matter halos in the non-Markov extension of EST via the Cholesky method which is discussed in previous section. (we assume the constant barrier). For a future work a more realistic collapse model must be studied). In Figure (\ref{fig:nM}) the number density of the dark matter halos is plotted with modified initial condition. In the lower panel of Figure (\ref{fig:nM}), we plot the mentioned ratio of the number density with respect to the standard Markov case. The figure shows  the excess in the power amplitude in scales of $k_*\simeq 2.72 h/Mpc$ ($M_* \simeq 10^{11}M_{\odot}$) causes an increase to excess in the mildly larger range ($M>M_*$) and also a decrease in the mass range of $M<10^{11} M_{\odot}$. This decrease in the number density of dark matter halos can naturally solve the TBTF problem and relax the tension in missing satellite problem. As discussed in the introduction the TBTF problem is also present in the field galaxies. This means that the tidal stripping mechanism which are introduced in galactic halos for solving this problem is not applicable to the field galaxies. However our proposal of modified initial condition is almost independent of the galaxies environment. \\
Another way to look at this proposal is via the conditional mass function. In Figure(\ref{fig:cond}), we plot the distribution of dark matter halo progenitors for a host halo with the mass of Milky way. The dark matter halo with the mass of $10^{12} M_{\odot}$ (Milky Way mass) is considered in present time ($z=0$) and the distribution of dark matter sub-halos is plotted in $z=1$. We see an excess over the mass scale $ M >10^{11} M_{\odot}$ and a decrease in number of the smaller dark matter halos. In this procedure, we us an approximation where the small scale dark matter halos merge more effectively to form larger halos. In other words, the larger merger rate for smaller halos causes to a deficiency in smaller dark matter halos and an excess for larger ones.

%%**************************************%%
%%**************************************%%
%%**************************************%%
\begin{figure}
	\includegraphics[width=\columnwidth]{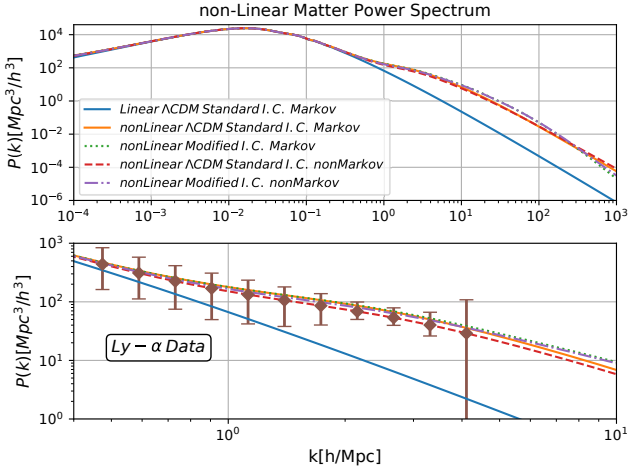}
    \caption{The nonlinear matter power spectrum for Markov and non-Markov case versus the wavenumber based on the halo model of dark matter. The lower panel is the zoomed in part of the non-linear power spectrum.} \label{fig:nl}
\end{figure}

%%**************************************%%
Now an  important issue which we should emphasis is that the non-linear power spectrum is well constrained by Ly-$\alpha$ observations. Accordingly the modification in the initial condition must not be in contradiction with the small scale non-linear power spectrum. %So by using equation (\ref{eq:1h2h}) and the number density of dark matter halos obtained from Cholesky method we derive the modified non-linear power spectrum.
In Figure(\ref{fig:nl}), we plot the halo-model nonlinear matter power spectrum, for the non-Markov and the modified case. In this figure, we show  relative 2$\sigma$ error bars of the Ly-$\alpha$  \cite{Zaroubi:2005xx} on the arbitrary chosen amplitude of the non-linear non-Markov power spectrum is consistent with our modified initial conditions. It should be noted that the normalized error bar with respect to power spectrum is used because the absolute  value of the Ly-$\alpha$ is obtained from direct inversion method by applying the redshift space distortion and non-linear effects, which is not considered in this work.
Finally summing up this section; the modified initial condition can be considered as promising candidate to solve the galactic scale problems of CDM paradigm, specially by relaxing the TBTF tension. In the last section, we will conclude and sketch the future prospects of this work.

%*****************************************************
%*****************************************************
%*****************************************************
%*****************************************************

\section{Conclusion and Future remarks}
\label{Sec:5}
{{The standard model of cosmology based on the cosmological constant and cold dark matter paradigm explains almost all the observational data from CMB to LSS.
However, there are couple of tensions for the standard model which may point to a new modified model beyond the standard one. A category of these tensions is related to the galactic scale challenges of CDM. Such as TBTF problem, missing satellite  and core-cusp problem. These challenges may indicate to a new modified initial condition or to a modified CDM model.  On the other hand the CMB and LSS observations constrain the initial power spectrum up to $\sim 1 \text{Mpc}$, accordingly a fundamental question can be raised; does the scale invariant power spectrum predicted by standard picture work to smaller scales or not?\\
In this work, we suggest a new idea to overcome the above mentioned problems.  We show that a simple deviation from standard inflationary paradigm (i.e. deviation from scale invariant prediction of  primordial perturbations) can be used as an alternative to the deviation from collisionless dark matter paradigm.
The Gaussian bump-like excess in initial power spectrum which can be considered as a natural outcome of inflationary models (many early universe models have an excess power in small scales, see the references in introduction) is proposed as a solution to this problem.}}
This modification change the number density distribution of dark matter halos in a nontrivial manner at the specific scale $k_* \simeq 3 \text{h/Mpc}$ where the modification in power spectrum is applied  (related to $M_* \simeq 10^{11}M_{\odot}$). We have a decrease in $M < M_*$ and increase $M >M_*$.  TBTF problem is related to the dark matter halos of mass $M<10^{11}M_{\odot}$. We show that this scale is related to $k*$. The solution to TBTF problem is considered to be the deficit of dark matter sub-halos due to this modification in initial power.\\

%This modification will change the number density of dark matter halos in a non-trivial manner. In a specific scale of $k_*$, where the modification in power spectrum is applied, we see a change in the distribution of the dark matter halos. We have a decrease in the number of halos in the mass scale $M < M_*$ and increase in larger mass scales. TBTF problem is related to the dark matter halos of mass $M<10^{11}M_{\odot}$. We show that this scale is related to $k* \sim 3 \text{h/Mpc}$. \\
In this work we use the non-Markov extension of the Excursion set theory via Cholesky method to study the modification. The non-Markov extension is necessary to use more realistic smoothing functions. We plot the number density distribution of dark matter halos using the standard $\Lambda$CDM model (and modified I. C. model) via cholesky non-Markov EST framework.
% In this direction, for the first time we plot the number density of dark matter halos using the standard model of $\Lambda$CDM via the Cholesky non-Markov EST method.
We also show that our modification to initial power spectrum is consistent with the non-linear power spectrum, which is well constrained by Ly-$\alpha$ data. We should note that the complexity of baryonic physics, the problem of the abundance matching (the relation connection between halo mass and stellar mass) has also a great importance. More sophisticated hydrodynamical simulations is needed to address these questions. However, semi analytical models which studied the deviations from the standard picture can be used as strong proposal for beyond standard model simulations. \\
We should note that our proposed method has a prediction as well  the number density of Milky way type galaxies must be larger than the standard case.
 In future works, we should consider some improvements. First a more realistic modification to the initial power spectrum emerged from viable inflationary models, second we should take into account a more realistic collapse model which leads to a moving barrier in EST context and third we have to consider a more realistic halo bias term.
As a final word, the future observations like large synoptic survey telescope (LSST), will probe the field galaxies beyond the Milky way and Andromeda. These new data samples will shed light on the missing satellite and TBTF problems. \\

\section*{Acknowledgements}

We are grateful to Ali Akbar Abolhasani, Shahram Khosravi, Aryan Rahimieh and Sohrab Rahvar for insightful comments and discussions.
We should also thank the anonymous referee for his/her valuable and insightful comments, which help us to improve the manuscript, specially in presenting the motivation of the work.
SB thanks the the Abdus Salam International Center of Theoretical Physics (ICTP) for a very kind hospitality, which initiation of  this work has been developed there.  SB is partially supported by Abdus Salam International Center of Theoretical Physics (ICTP) under the junior associateship  scheme during this work. This research is supported Sharif University of Technology Office of Vice President for Research under Grant No. G960202  \\ \\

%%%%%%%%%%%%%%%%%%%%%%%%%%%%%%%%%%%%%%%%%%%%%%%%%%
%%%%%%%%%%%%%%%%%%%% REFERENCES %%%%%%%%%%%%%%%%%%

% The best way to enter references is to use BibTeX:
%\bibliographystyle{mnras}
%\bibliography{example} % if your bibtex file is called example.bib
% Alternatively you could enter them by hand, like this:
% This method is tedious and prone to error if you have lots of references

%%%%%%%%%%%%%%%%%%%%%%%%%%%%%%%%%%%%%%%%%%%%%%%%%%
%%%%%%%%%%%%%%%%% APPENDICES %%%%%%%%%%%%%%%%%%%%%

%\appendix
%\section{Filament length in ellipsoidal collapse} \label{app:length}

% Don't change these lines
\bsp	% typesetting comment
\label{lastpage}
\end{document}